\documentclass[prd,twocolumn,tightenlines,showpacs,amsmath,amssymb]{revtex4-1}

\usepackage{amssymb}
\usepackage{multirow}
\usepackage{graphicx,epsfig}
\usepackage{dcolumn}
\usepackage{bm}
\usepackage[dvipdfm,CJKbookmarks=true,unicode,colorlinks,linkcolor=blue,anchorcolor=blue,citecolor=blue,pdfborder={0 0 0}]{hyperref}
\usepackage{amsfonts}
\usepackage{keyval,graphicx}
\usepackage{textcomp,wasysym}

\begin{document}
\newcommand{\cinst}[2]{$^{\mathrm{#1}}$~#2\par}
\newcommand{\crefi}[1]{$^{\mathrm{#1}}$}
\newcommand{\crefii}[2]{$^{\mathrm{#1,#2}}$}
\newcommand{\crefiii}[3]{$^{\mathrm{#1,#2,#3}}$}
\newcommand{\HRule}{\rule{0.5\linewidth}{0.5mm}}
\newcommand{\br}[1]{\mathcal{B}#1}
\newcommand{\el}[1]{\mathcal{L}#1}
\newcommand{\ef}[1]{\mathcal{F}#1}



\title{\boldmath Study of $J/\psi\to \eta\phi\pi^+\pi^-$ at BESIII}

\author{
  {\small M.~Ablikim$^{1}$, M.~N.~Achasov$^{8,a}$, X.~C.~Ai$^{1}$,
      O.~Albayrak$^{4}$, M.~Albrecht$^{3}$, D.~J.~Ambrose$^{43}$,
      A.~Amoroso$^{47A,47C}$, F.~F.~An$^{1}$, Q.~An$^{44}$,
      J.~Z.~Bai$^{1}$
% ...
      \\
      \vspace{0.2cm}
      (BESIII Collaboration)\\
      \vspace{0.2cm} {\it
        $^{1}$ Institute of High Energy Physics, Beijing 100049, People's Republic of China\\
% ...
$^{3}$ Bochum Ruhr-University, D-44780 Bochum, Germany\\
        $^{4}$ Carnegie Mellon University, Pittsburgh, Pennsylvania 15213, USA\\
% ...
$^{8}$ G.I. Budker Institute of Nuclear Physics SB RAS (BINP), Novosibirsk 630090, Russia\\
% ...
$^{43}$ University of Rochester, Rochester, New York 14627, USA\\
        $^{44}$ University of Science and Technology of China, Hefei 230026, People's Republic of China\\
% ...
$^{47}$ (A)University of Turin, I-10125, Turin, Italy; (B)University of Eastern Piedmont, I-15121, Alessandria, Italy; (C)INFN, I-10125, Turin, Italy\\
% ...
$^{a}$ Also at the Novosibirsk State University, Novosibirsk, 630090, Russia\\
% ...
     \vspace{0.8cm}
}}
}
\begin{abstract}
Based on a sample of $2.25\times 10^{8} J/\psi$ events taken with the
BESIII detector at the BEPCII collider, we present the results of a
study of the decay $J/\psi\to \eta \phi\pi^{+}\pi^{-}$. The $Y(2175)$ resonance
is observed in the invariant mass spectrum of $\phi f_{0}(980)$ with a
statistical significance of greater than $10\sigma$. The corresponding mass and
width are determined to be $M=2200\pm 6 \mathrm{(stat.)} \pm
5\mathrm{(syst.)}~\mathrm{MeV}/c^{2}$ and $\Gamma=104\pm
15\mathrm{(stat.)}\pm 15\mathrm{(syst.)}$~MeV, respectively, and the
product branching fraction is measured to be
$\mathcal{B}(J/\psi\to\eta Y(2175)$, $Y(2175)\to \phi f_{0}(980)$,
$f_{0}(980)\to \pi^{+}\pi^{-})= (1.20\pm 0.14\mathrm{(stat.)}\pm 0.37
\mathrm{(syst.)})\times 10^{-4}$. The results are consistent within
errors with those of previous experiments. We also measure the
branching fraction of $J/\psi\to \phi f_1(1285)$ with $f_1(1285)\to
\eta\pi^{+}\pi^{-}$ and set upper limits on the branching fractions
for $J/\psi\to \phi\eta(1405)$/$\phi X(1835)$/$\phi X(1870)$ with
$\eta(1405)$/$X(1835)$/$X(1870)\to \eta\pi^{+}\pi^{-}$ at the 90\%
confidence level.

\end{abstract}


\maketitle

\section{Introduction}

The $Y(2175)$, also referred to as the $\phi(2170)$ by the Particle
Data Group~(PDG 2014)~\cite{PDG}, was first observed by the BABAR
experiment~\cite{aaaa_babar} in the $e^+e^-\rightarrow\gamma_{ISR}\phi
f_0(980)$ initial-state-radiation~(ISR) process. It was later
confirmed by the BESII experiment in $J/\psi\rightarrow \eta\phi
f_0(980)$ decays~\cite{aaab_wanx} and via the same ISR process by the
BELLE~\cite{aaba_belle} and BABAR experiments~\cite{babar_y2175} with
increased statistics. Since the $Y(2175)$ resonance is produced via
ISR in $e^{+}e^{-}$ collisions, it is known to have
$J^{PC}=1^{--}$.   This observation stimulated the
speculation that the $Y(2175)$ may be an $s$-quark counterpart to the
$Y(4260)$~\cite{babay4260,bn978}, since both are produced in
$e^{+}e^{-}$ annihilation and exhibit similar decay patterns. Like for
the $Y(4260)$, a number of different interpretations have been proposed
for the $Y(2175)$ with predicted masses that are consistent, within
errors, with the experimental measurements. These include: an
$s\overline{s}$-gluon hybrid~\cite{Yml1}; an excited $\phi$
state~\cite{Yml2}; a tetraquark state~\cite{tetraq}; a $\Lambda
\overline{\Lambda}$ bound state~\cite{EK15,CFQ}; or an ordinary $\phi
f_{0}(980)$ resonance produced by interactions between the final state
particles~\cite{27mev}.

A recent review~\cite{zhusl} discusses the
basic problem of the large expected decay widths into two mesons,
which contradicts experimental observations. Around the mass of the
$Y(2175)$, there are two conventional $1^{--}$ $s\bar s$ states in the
quark model, $2 ^3D_1$ and $3 ^3S_1$. According to Ref.~\cite{barnes10},
the width of the $3^3S_1$ $s\bar s$ state is expected to be about 380 MeV. The total
width of the $2 ^3D_1$ state from both $^3P_0$ and flux tube model is
expected to be around $(150\sim 250)$ MeV \cite{Yml2}. However, the predictions from
these strong decay models sometimes deviate from the experimentally found
width by a factor of two or three. For comparison, the widths of the
$3^3S_1$ and $2 ^3D_1$ charmonium are less than 110 MeV
\cite{wmy}. Fortunately, the characteristic decay modes of Y(2175) as
either a hybrid or $s\bar s$ state are quite different, which may be
used to distinguish the hybrid and $s\bar s$ schemes. The possibility
of $Y(2175)$ arising from $S$-wave threshold effects is not excluded.  As
of now, none of these interpretations have been either established or
ruled out by experiment.  The confirmation and study of the $Y(2175)$
in $J/\psi\to\eta\phi\pi^{+}\pi^{-}$ with a large data sample is
necessary for clarifying its nature.

The $J/\psi\rightarrow\eta\phi\pi^+\pi^-$ decay also offers a
unique opportunity to investigate the properties of the $f_1(1285)$,
the $\eta(1295)$, and the $\eta(1405)/\eta(1475)$ resonances. The
$f_1(1285)$ is usually considered to be a member of the axial vector
meson nonet, but the interpretation of the $\eta(1295)$ is less
clear. Both the $f_1(1285)$ and the $\eta(1295)$ were seen in fixed
target experiments, but the $\eta(1295)$ was not evident in central
production, in $\gamma\gamma$ collisions, or in $J/\psi$
decays. Therefore it has been speculated that either the $f_1(1285)$,
at least in some cases, contains an $\eta(1295)$
component~\cite{both}, or that the $\eta(1295)$ does not exist. The
$\eta(1405)/\eta(1475)$ pseudoscalar was once regarded as a glueball
candidate since it is copiously produced in $J/\psi$ radiative
decays~\cite{jpsirad} and there was only an upper limit from
$\gamma\gamma$ collisions~\cite{ggc1}. But this viewpoint changed when
the $\eta(1405)/\eta(1475)$ was also observed in untagged
$\gamma\gamma$ collisions~\cite{L3} and in $J/\psi$ hadronic decays.

In addition, two interesting resonances, the $X(1835)$ and the
$X(1870)$, were observed in
$J/\psi\rightarrow\gamma\pi^+\pi^-\eta^\prime$~\cite{X1835_pipietap1,X1835_pipietap2}
and $J/\psi\rightarrow\omega\pi^+\pi^-\eta$~\cite{liuk},
respectively. The $X(1835)$, in particular, inspired many possible
theoretical interpretations, including a $p\overline{p}$ bound
state~\cite{ppbs1,ppbs2}, a glueball~\cite{X1835g1,X1835g2,X1835g3},
and final state interactions~(FSI) between a proton and
antiproton~\cite{fsi1,fsi2,fsi3}. To better understand the properties
of these two resonances, one needs to further study their production
in different $J/\psi$ decay modes. For example, the search for them in
the $\eta\pi^{+}\pi^{-}$ mass spectrum recoiling against the $\phi$ in
$J/\psi$ decays would be rather interesting for clarifying their
nature.

In this paper, we present a study of the decay
$J/\psi\rightarrow\eta\phi\pi^+\pi^-$ with $\eta\to\gamma\gamma$ and
$\phi\to K^{+}K^{-}$ decay modes using a sample of $2.25\times10^{8}$
$J/\psi$ events collected with the Beijing Spectrometer~(BESIII)
located at the Beijing Electron-Positron
Collider~(BEPCII)~\cite{BEPCII}. The mass and width of the $Y(2175)$,
as well as its production rate, are measured. In addition, the
production rates of the $f_1(1285)$, the $\eta(1405)/\eta(1475)$, the
$X(1835)$, and the $X(1870)$ in $J/\psi$ hadronic decays associated
with a $\phi$ meson are investigated.

\section{Detector and Monte Carlo simulation}

The BESIII detector is a magnetic spectrometer~\cite{BEPCII} located
at BEPCII, which is a double-ring $e^+ e^-$ collider with a design
peak luminosity of $10^{33}$ cm$^{-2}$ s$^{-1}$ at a center-of-mass
energy of 3.773 GeV. The cylindrical core of the BESIII detector
consists of a helium-based main drift chamber (MDC), a plastic
scintillator time-of-flight system (TOF), and a CsI (Tl)
electromagnetic calorimeter (EMC), which are all enclosed in a
superconducting solenoidal magnet providing a 1.0 T magnetic field.
The solenoid is supported by an octagonal flux-return yoke with
modules of resistive plate muon counters interleaved with steel. The
acceptance for charged particles and photons is 93\% of the full 4$\pi$
solid angle. The momentum resolution for a charged particle at 1 GeV/$c$
is 0.5\%, and the ionization energy loss per unit path-length
($dE/dx$) resolution is 6\%. The EMC measures photon energies with a
resolution of 2.5\% (5\%) at 1 GeV in the barrel (end-caps). The time
resolution for the TOF is 80 ps in the barrel and 110 ps in the
end-caps.

The GEANT-based simulation software BOOST~\cite{MC1} is used to
simulate the desired Monte Carlo~(MC) samples. An inclusive $J/\psi$
MC sample is used to estimate the backgrounds. The production of the
$J/\psi$ resonance is simulated by the MC event generator
KKMC~\cite{KKMC1,KKMC2}, while the decays are generated by
BesEvtGen~\cite{besevtgen1,besevtgen2,evtgen} for known decay modes with
branching fractions set at the PDG~\cite{PDG} world average values,
and by the Lund-Charm model~\cite{lundcharm} for the remaining unknown
decays.

In this analysis, a signal MC sample for the process $J/\psi\to\eta
Y(2175)$, $Y(2175)\to\phi f_{0}(980)$ and $f_{0}(980)\to\pi^+\pi^-$,
is generated to optimize the selection criteria and determine the
detection efficiency. Since the $J^{PC}$ of the $Y(2175)$ is $1^{--}$,
a $P$-wave orbital angular momentum is used for the $\eta-Y(2175)$
system, while $S$-wave is used for the $\phi-f_{0}(980)$ and
$\pi^+-\pi^-$ systems. The shape of the $f_{0}(980)$ is parameterized
with the $\rm Flatt\acute{e}$ formula~\cite{Flatte}, and the
corresponding parameters are taken from the measurement of
BESII~\cite{phipipi}. For the signal MC sample of $J/\psi\to\phi
\eta(1405)/ f_1(1285)$, the angular distributions are also considered
in the simulation.

\section{Event selection}

To select candidate events of the process
$J/\psi\to\eta\phi\pi^{+}\pi^{-}$ with $\phi\to K^{+}K^{-}$ and
$\eta\to \gamma \gamma$, the following criteria are imposed on the
data and MC samples.

We select charged tracks in the MDC within the polar angle range
$|\cos\theta|<0.93$ and require that the points of closest approach to
the beam line be within $\pm 20$~cm of the interaction point in the
beam direction and within $2$~cm in the plane perpendicular to the
beam. The TOF and $dE/dx$ information are combined to form particle
identification~(PID) confidence levels for the $\pi$, $K$, $p$
hypotheses, and each track is assigned to the particle type
corresponding to the hypothesis with the highest confidence level. Two
kaon and two pion particles with opposite charges are required.

Photon candidates are reconstructed by clustering signals in EMC
crystals. The energy deposited in nearby TOF counters is included to
improve the photon reconstruction efficiency and the photon energy
resolution. At least two photon candidates are selected, the minimum energy of
which are required to be $25$~MeV for barrel showers ($|\cos\theta| <
0.80$) and $50$~MeV for endcap showers ($0.86<|\cos\theta|<0.92$). To
exclude showers due to the bremsstrahlung of charged particles, the
angle between the nearest charged track and the shower must be greater
than $10^{\circ}$. EMC cluster timing requirements are applied to
suppress electronic noise and energy deposits unrelated to the event.

A four-constraint kinematic fit using energy-momentum conservation is
performed to the $J/\psi\to K^{+}K^{-}\pi^{+}\pi^{-}\gamma\gamma$
hypothesis. All combinations of two photons are tried and the one with
the smallest $\chi^{2}_{4C}$ value is retained. To further suppress
background, $\chi^2_{4C}<200$ is required.

\begin{figure*}[htbp]
   \includegraphics[width=0.33\textwidth]{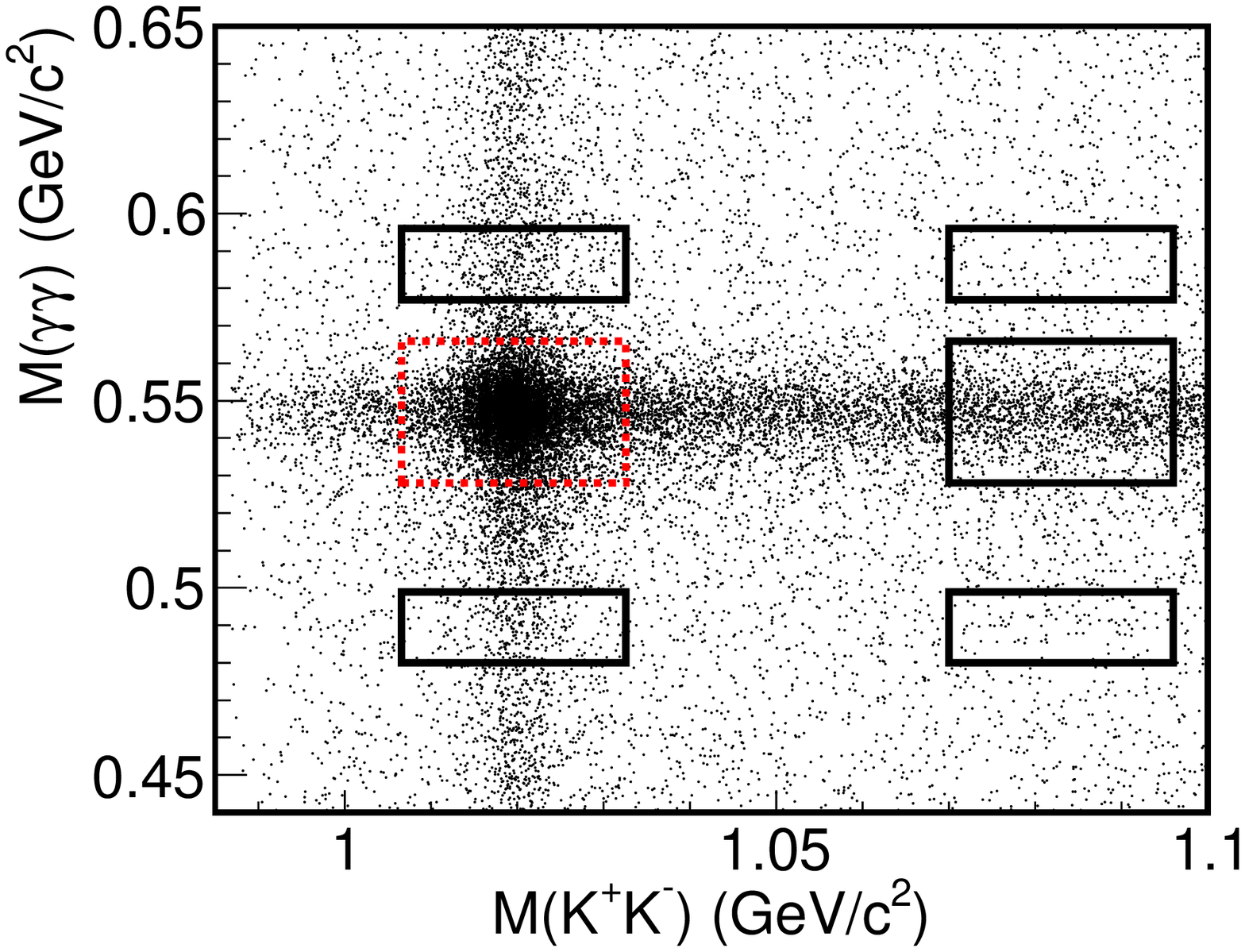}
   \put(-60,100){(a)}
   \includegraphics[width=0.33\textwidth]{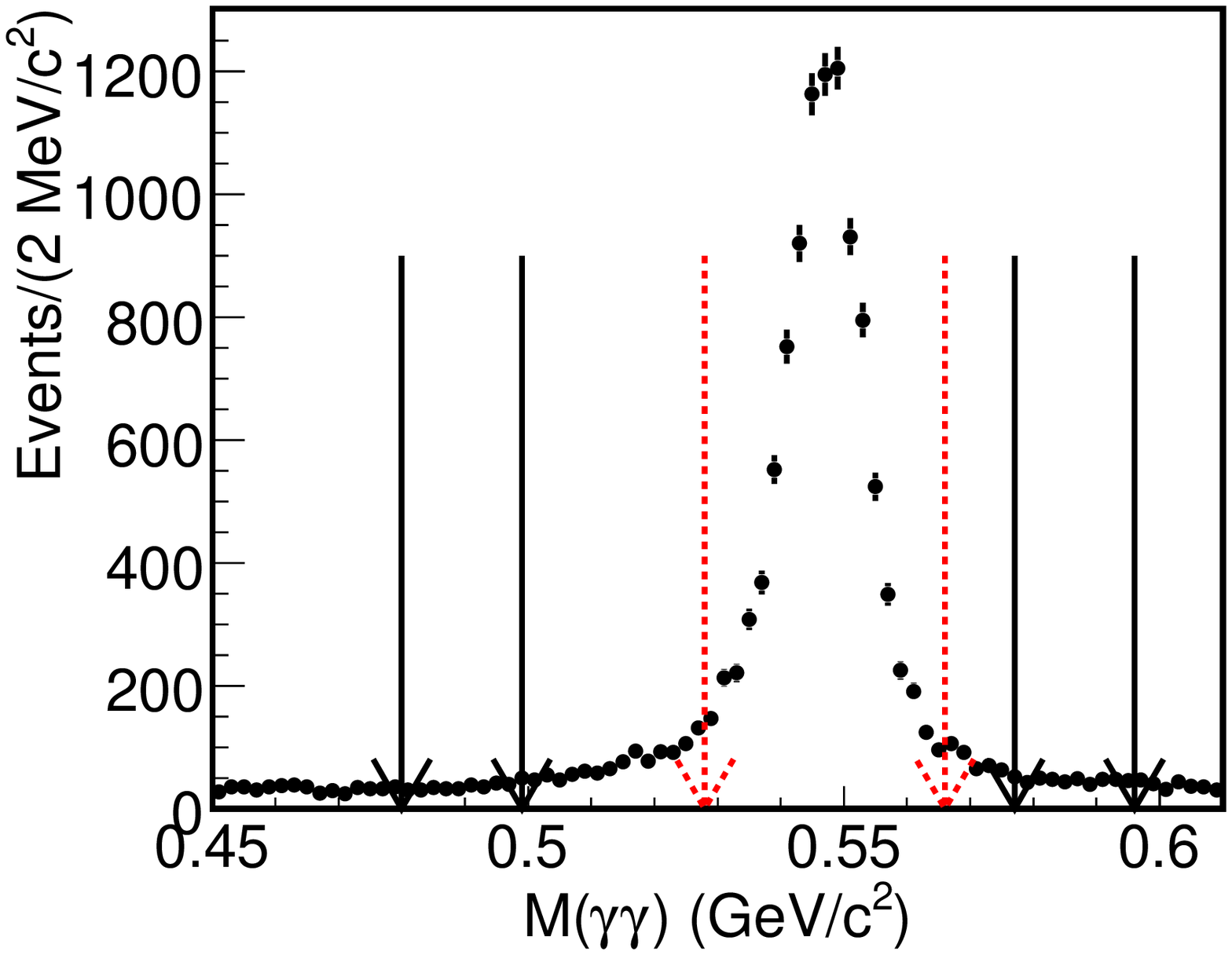}
   \put(-60,100){(b)}
   \includegraphics[width=0.33\textwidth]{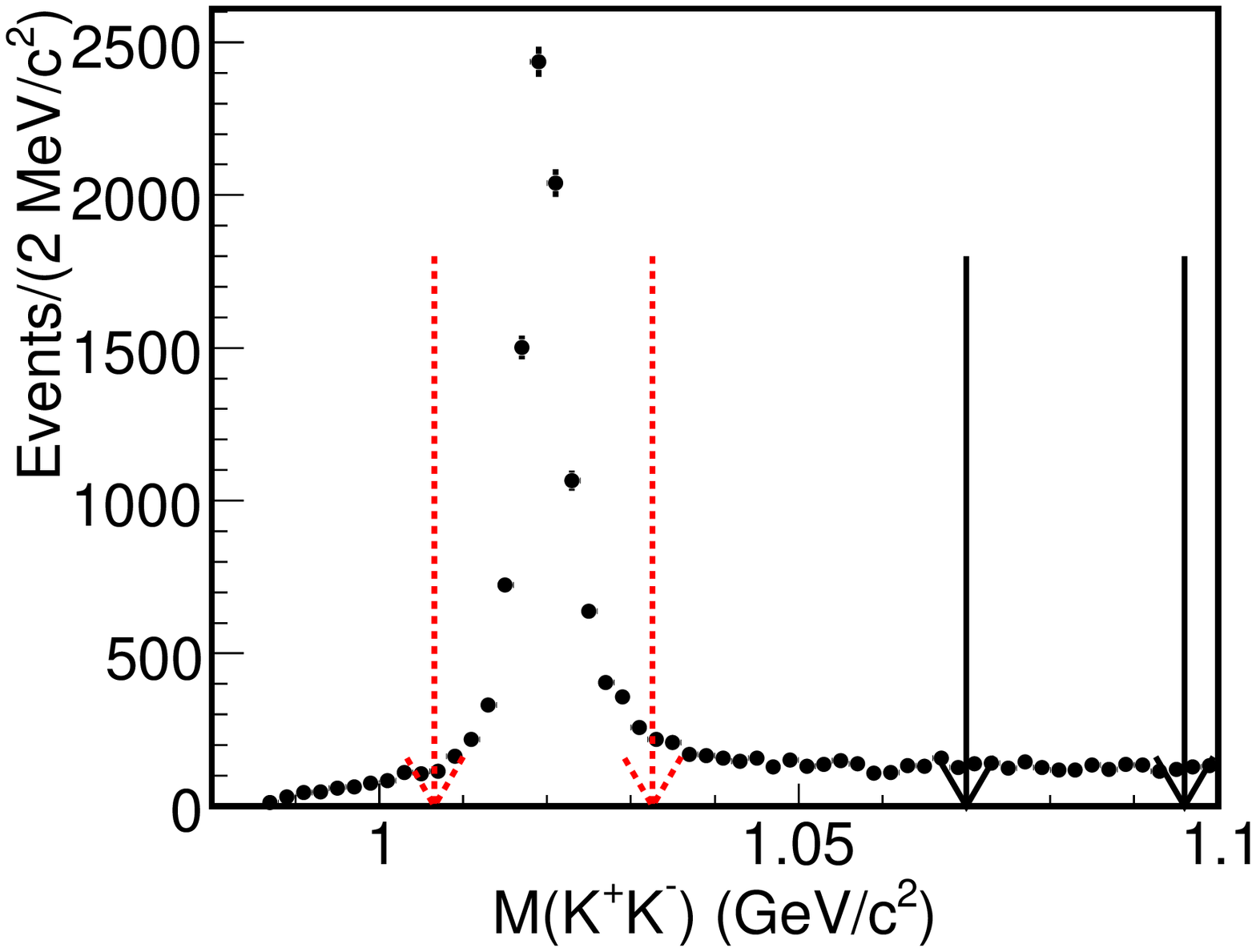}
   \put(-60,100){(c)}
   \caption{ (a)~Scatter plot of $M(\gamma\gamma)$ versus
     $M(K^{+}K^{-})$. The boxes with the dotted and solid lines show
     the $\eta$ and $\phi$ signal and sidebands regions,
     respectively. (b)~The $\gamma\gamma$ invariant mass spectrum for
     events with the $K^{+}K^{-}$ invariant mass in the $\phi$ signal
     region. (c)~The $K^{+}K^{-}$ invariant mass spectrum for events
     with the $\gamma\gamma$ invariant mass in the $\eta$ signal
     region. In plots~(b) and~(c), the dotted arrows show the signal
     regions and the solid lines show the sideband regions, which are
     described in the text. \label{ggvskk}}
\end{figure*}

After the above selection process, a scatter plot of the invariant
mass of the $\gamma\gamma$ system~($M(\gamma\gamma)$) versus the
invariant mass of the $K^{+}K^{-}$ system~($M(K^+K^-)$) in data is shown in
Fig.\ref{ggvskk}(a), where the events concentrated in the region
indicated by the dotted-line box correspond to the $J/\psi\to\eta\phi
\pi^{+}\pi^{-}$ signal. The $\phi$ and $\eta$ signal regions are
defined as $|M(K^{+}K^{-})-M_{\phi}|<0.013$~GeV$/c^{2}$ and
$|M(\gamma\gamma)-M_{\eta}|<0.019$~GeV$/c^{2}$, where $M_{\phi}$ and
$M_{\eta}$ are world average values of the $\phi$ and $\eta$ masses,
respectively. Fig.\ref{ggvskk}(b) and~(c) show the $\gamma\gamma$ and
$K^{+}K^{-}$ invariant mass distributions for events with a $K^{+}K^{-}$
invariant mass within the $\phi$ signal region and a $\gamma\gamma$
invariant mass within the $\eta$ signal region, respectively. Both
$\eta$ and $\phi$ signals are clearly seen with very low background
levels.

\section{\boldmath Measurement of $J/\psi\to\eta Y(2175)$ with $Y(2175)\to\phi f_{0}(980)$ and $f_{0}(980)\to\pi^{+}\pi^{-}$} \label{jpsi_channel}

\begin{figure*}[htbp]
   \includegraphics[width=6cm]{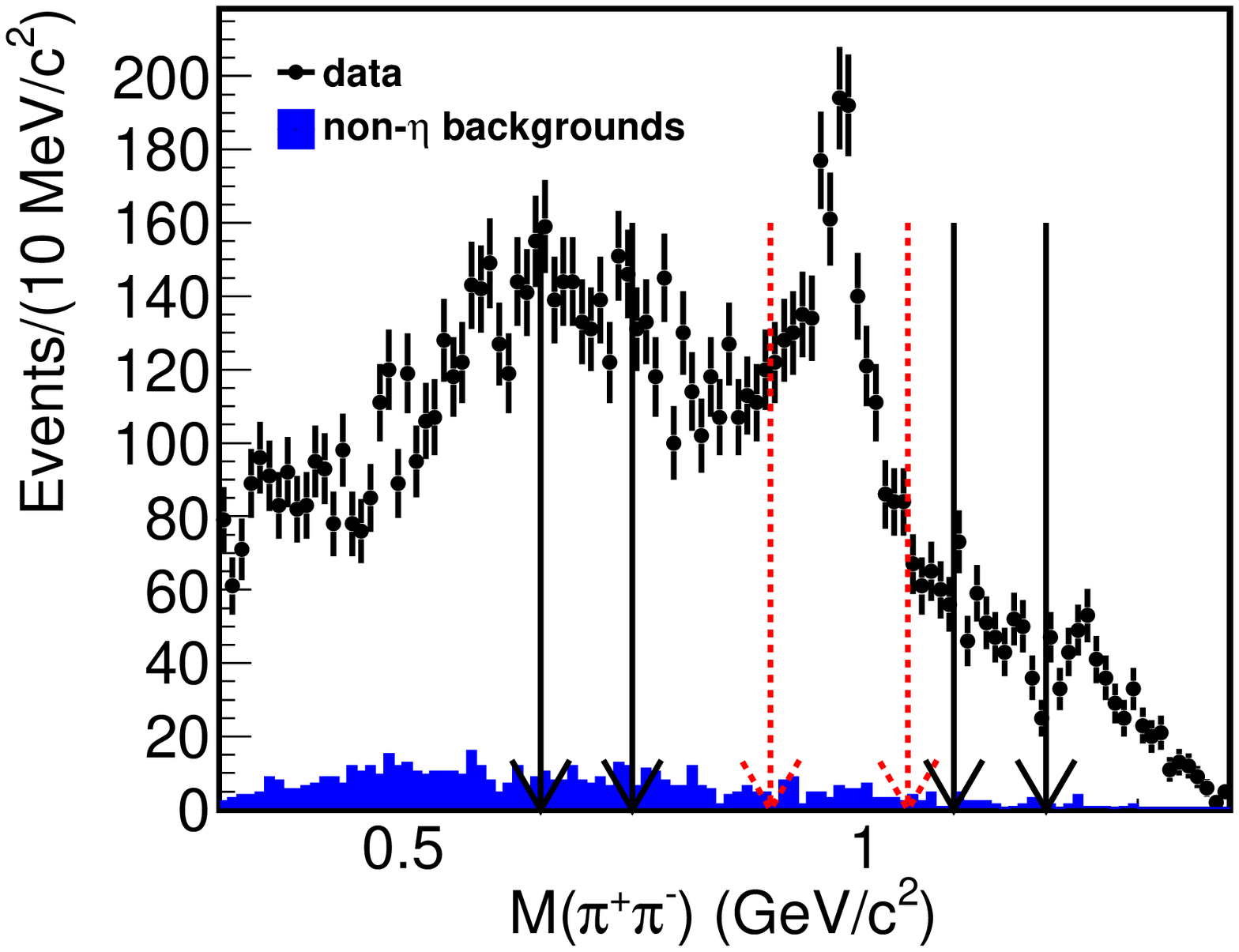}
              \put(-60,100){(a)}
   \includegraphics[width=6cm]{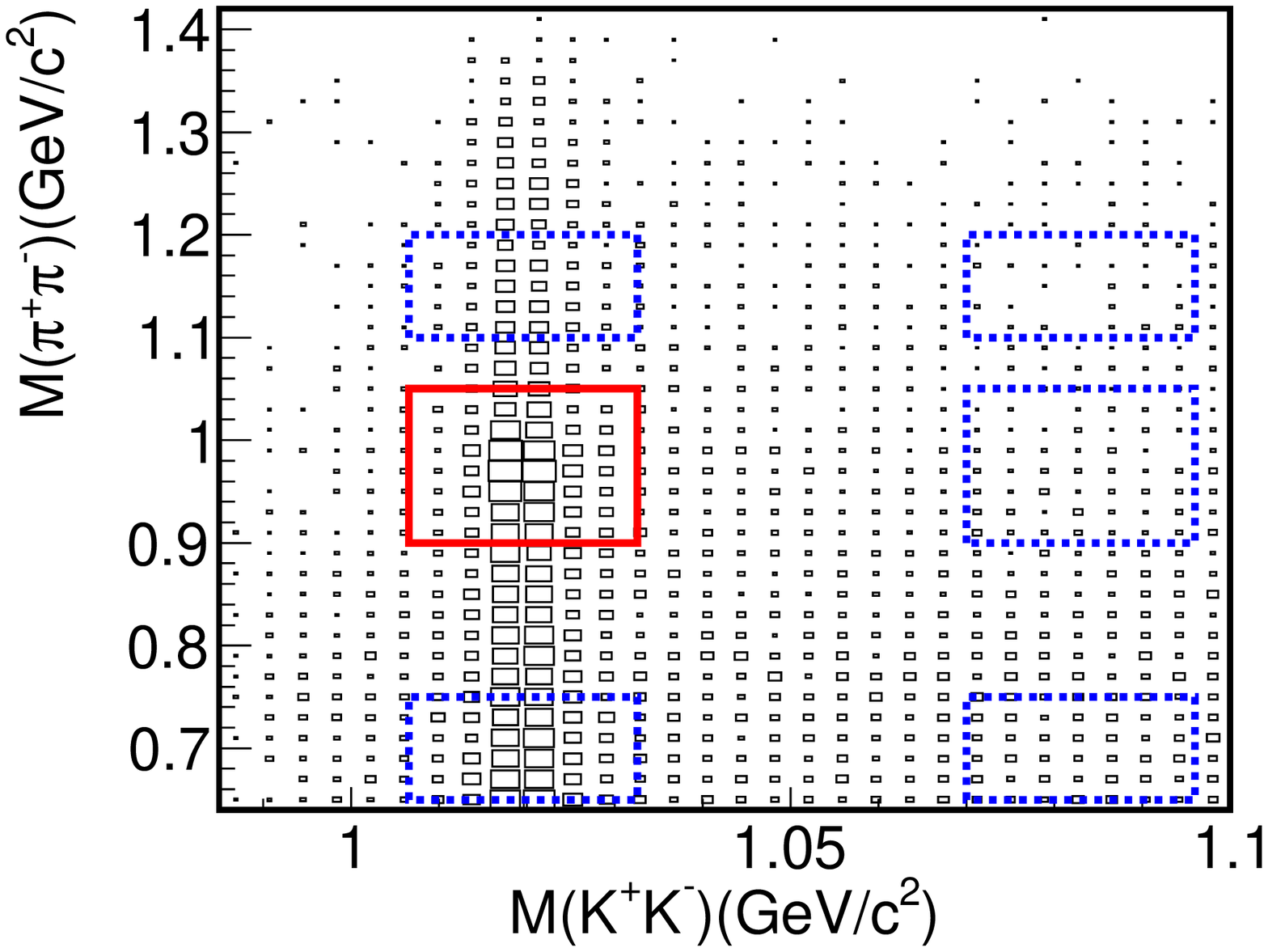}
              \put(-60,100){(b)}

   \includegraphics[width=6cm]{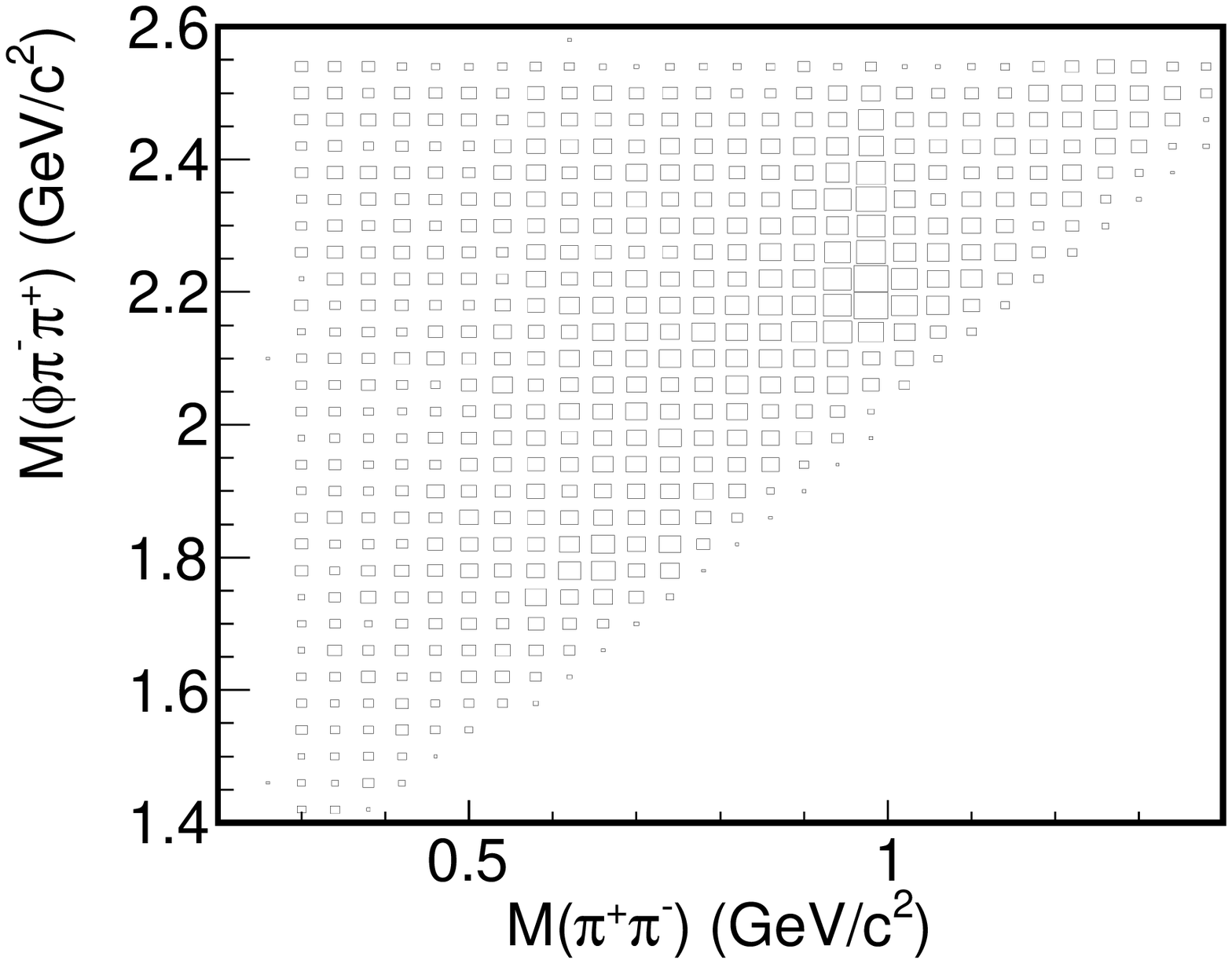}
       \put(-60,60){(c)}
   \includegraphics[width=6cm]{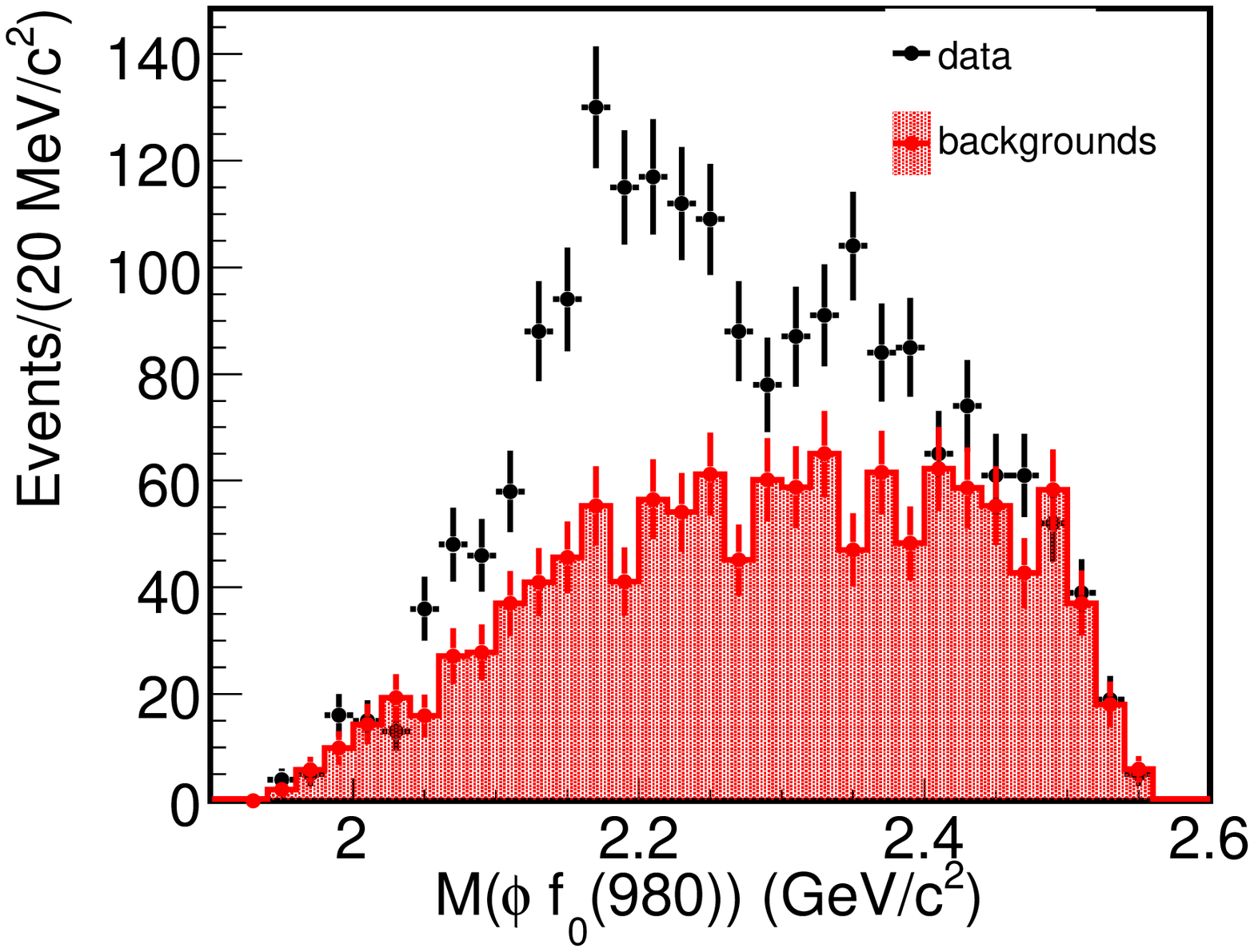}
       \put(-120,100){(d)}
       \caption{
     (a) The $\pi^{+}\pi^{-}$ invariant mass spectrum. The shaded histogram shows the non-$\eta$ background estimated with $\eta$ sideband region; the dotted and solid arrows denote the $f_{0}(980)$ signal and sideband regions, respectively.
     (b) The scatter plot of $M(\pi^{+}\pi^{-})$ versus $M(K^{+}K^{-})$. The solid box shows the signal region, and the dotted boxes show the sideband regions of $\phi$ and $f_{0}$.
     (c) The scatter plot of $M(\phi\pi^{+}\pi^{-})$ versus $M(\pi^{+}\pi^{-})$.
     (d) The $\phi\pi^{+}\pi^{-}$ invariant mass distribution after
     imposing the $f_{0}(980)$ signal mass window requirement. The
     shaded histogram shows the background distribution estimated with
     the sideband method described in the text.
   \label{c1}}
\end{figure*}

With the above requirements on the $\eta$ and $\phi$ candidate masses, the
$\pi^{+}\pi^{-}$ invariant mass distribution is shown in
Fig.~\ref{c1}(a).  A clear $f_{0}(980)$ signal is visible.  The
non-$\phi$ and/or non-$\eta$ backgrounds are estimated with the events
in the $\eta-\phi$ sideband regions, shown as the shaded histogram in
Fig.~\ref{c1}(a). The $\eta$ sideband is defined by
$0.480$~GeV$/c^2<M(\gamma\gamma)<0.499$~GeV$/c^{2}$ or
$0.577$~GeV/$c^2<M(\gamma\gamma)<0.596$~GeV$/c^{2}$, and the $\phi$
sideband is defined by
$1.070$~GeV$/c^2<M(K^{+}K^{-})<1.096$~GeV$/c^2$. Using a mass
requirement of $0.90$~GeV$/c^{2}<M(\pi^{+}\pi^{-})<1.05$~GeV$/c^{2}$
to select the $f_0(980)$ signal, the invariant mass distribution of
$\phi f_0(980)$ is shown in Fig.~\ref{c1}(d), where a broad structure
around $2.2$~GeV$/c^2$ is evident. Figure~\ref{c1}(c) shows a
two-dimensional histogram of $M(\phi\pi^{+}\pi^{-})$ versus
$M(\pi^{+}\pi^{-})$. A cluster of events populating the $Y(2175)$ and
$f_{0}(980)$ signal regions is observed, which corresponds to the
decay of $Y(2175)\to\phi f_{0}(980)$ with
$f_{0}(980)\to\pi^{+}\pi^{-}$.

Since the contribution from non-$\eta$ background events in the $f_0(980)$
mass region is small and can be neglected, the two-dimensional
$\phi$-$f_0(980)$ sidebands are used to estimate the background
events in this analysis.  With the $\eta$ mass requirement applied, the
non-$\phi$ and/or non-$f_0(980)$ events are estimated by the
weighted sums of horizontal and vertical sidebands, with the entries
in the diagonal side bands subtracted to compensate for the double
counting of background components. The definition of the
two-dimensional side bands is illustrated in Fig.~\ref{c1}(b). The
weighting factors for the events in the horizontal, vertical and the
diagonal side bands are measured to be 0, and 0.66, -0.085
respectively, which are determined from the results of a
two-dimensional fit to the mass spectrum of $M(K^{+}K^{-})$ versus
$M(\pi^{+}\pi^{-})$. No signal of $f_{0}(980)$ is evident in
non-$\phi$ processes as shown in the scatter plot of $M(\pi^{+}\pi^{-})$
versus $M(K^{+}K^{-})$.  Hence, the
weighting factor for the events in the horizontal side band is zero,
and the non-$\phi$ events in the horizontal side band are not used in
the background estimation.  The two-dimensional Probability
Density Functions (PDFs) for $J/\psi\to\eta\phi f_0(980)$, $\phi$ but
non-$f_0(980)$, non-$\phi$ and non-$f_0(980)$ processes are
constructed by the product of one-dimensional functions, where the
resonant peaks are parameterized by Breit-Wigner functions (for
$\phi$) and a shape taken from simulation (for $f_0(980)$), and the non-resonant parts are
described by polynomials with coefficients left free in the fit. To account for the difference of
the background shape between the signal region and side bands due to
the varying phase space, the obtained background mass distribution is
multiplied by a correction curve determined from an MC sample of 1
million events of the phase space processes $J/\psi\to\eta \phi
\pi^{+}\pi^{-}$. The estimated $K^{+}K^{-}\pi^{+}\pi^{-}$ invariant
mass distribution for the total non-$\phi$ or non-$f_0(980)$
components is shown by the shaded histogram in Fig.~\ref{c1}(d).
No evident $Y(2175)$ signal is observed.

To extract the yield of $Y(2175)$, an unbinned maximum likelihood fit
to the $\phi f_{0}(980)$ invariant mass is performed. The $Y(2175)$
signal, the direct three-body decay of $J/\psi\to\eta\phi f_{0}(980)$,
and the background from the above estimation shown as the shaded
histogram in Fig.~\ref{c1}(b) are included in the fit. With the
assumption of no interference between the $Y(2175)$ signal and the
direct three-body decay of $J/\psi\to\eta\phi f_{0}(980)$, the
probability density function~(PDF) can be written as
\begin {equation}
\label{eqfit0}
\epsilon(m)\times(G\otimes |A(m)|^2)+ A(J/\psi\to\eta\phi f_0)  + \rm BKG,
\end {equation}
where $A(m)=\frac{P_{J\to\eta Y}^{l_{1}}P_{Y\to\phi
    f_{0}}^{l_{2}}}{m^2-M_{0}^2+iM_{0}\Gamma_{0}}$ is a Breit-Wigner
function representing the $Y(2175)$ signal shape, taking into account
the phase space factor of a two-body decay.  $M_{0}$ and $\Gamma_{0}$
are left free in the fit. $P_{J\to\eta Y}$ and $P_{Y\to\phi f_{0}}$
denote the momentum of the $\eta$ in the rest frame of the $J/\psi$
and that of the $\phi$ in the rest frame of the $Y(2175)$,
respectively. $l_{1}$ and $l_{2}$, which label the relative orbital
angular momenta of the $\eta-Y(2175)$ and $\phi-f_{0}(980)$ systems,
are set to be $1$ and $0$ in the fit, respectively. $G$ is a Gaussian
function representing the mass resolution, and the corresponding
parameters are taken from MC simulation. $\epsilon(m)$, the detection
efficiency as a function of the $\phi f_{0}(980)$ invariant mass, is
also obtained from MC simulation. $A(J/\psi\to\eta\phi f_0)$
represents the component of the direct decay of $J/\psi\to\eta\phi
f_{0}(980)$ with the shape derived from the phase space MC sample. Finally,
$\rm BKG$ refers to the background component estimated from the
two-dimensional weighted sideband method.

\begin{figure}[htbp]
   \includegraphics[width=\columnwidth]{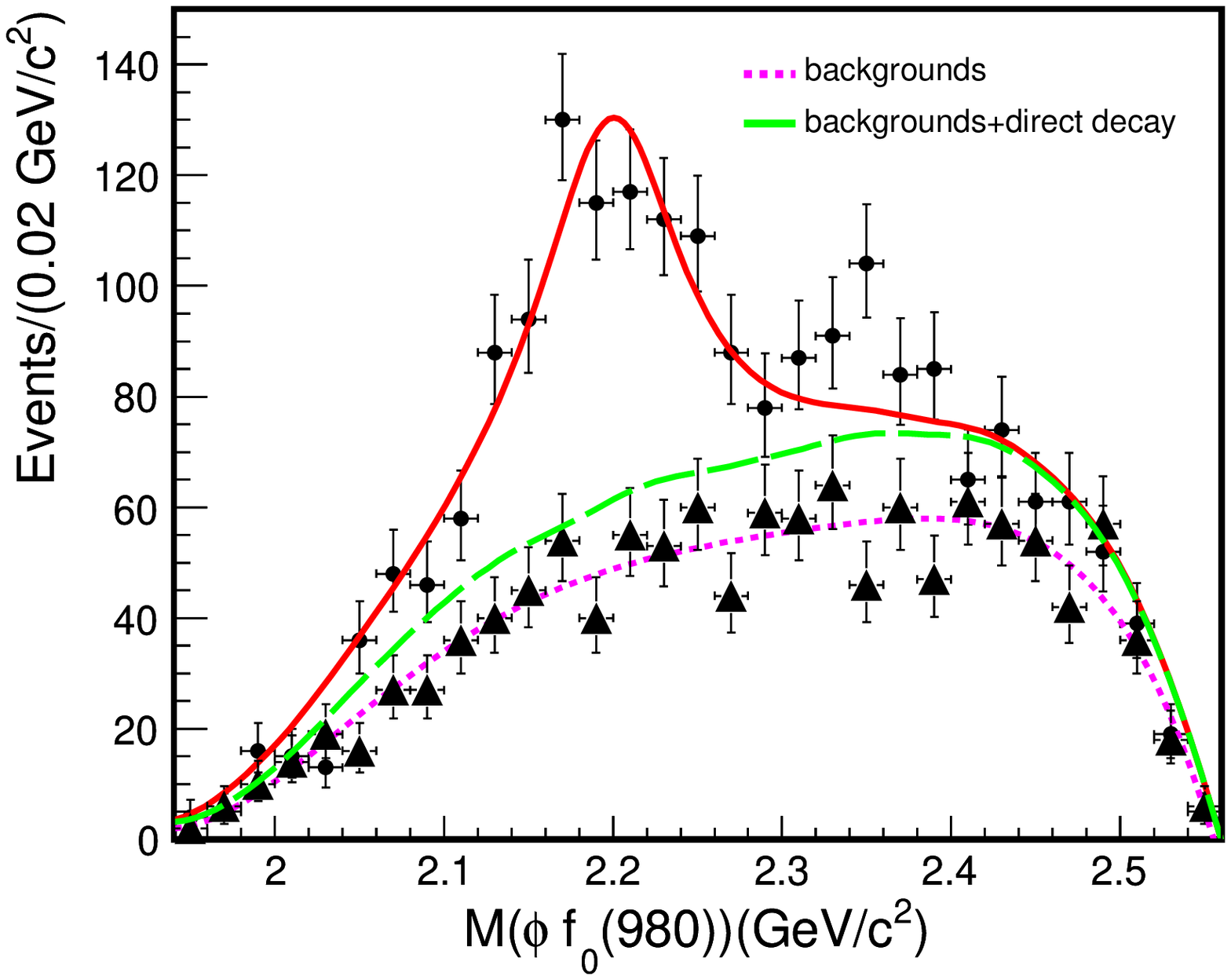}
   \caption{Result of the fit to the $\phi f_0(980)$
       invariant mass distribution described in the text. The
     circular dots with error bars show the distribution in the signal
     region; the triangular dots with error bars show the backgrounds
     estimated using sideband regions; the solid curve shows the
     overall fit projection; the dotted curve shows the fit for the
     backgrounds; and the dashed curve is for the sum of the direct
     decay of $J/\psi\to\eta\phi f_0$ and backgrounds.
     \label{fitY2}}
\end{figure}

Figure~\ref{fitY2} shows the results of the fit, where the circular
dots with error bars show the distribution for the signal and the
triangular dots with error bars are for the backgrounds estimated by
the sidebands. The solid curve is the overall fit projection, the dotted
curve the fit for the backgrounds, and the dashed curve for the sum of
the direct decay of $J/\psi\to\eta\phi f_0$ and backgrounds. The mass
and width of the $Y(2175)$ are determined to be $M=2200\pm
6$~MeV/c$^2$ and $\Gamma=104\pm 15$~MeV, respectively. The fit yields
$471\pm 54$ $Y(2175)$ events with a statistical significance of
greater than 10$\sigma$, which is determined by the change of the log-likelihood
value and the number of degree of freedom in the fit with and without
the $Y(2175)$ signal.  Taking into account the detection efficiency,
$(9.10\pm0.01)\%$, obtained from MC simulation, the product branching fraction
is
\begin{align*}
 \mathcal{B}(&J/\psi\to\eta Y(2175), Y(2175)\to\phi f_0(980), \\
 & f_0(980)\to\pi^{+}\pi^{-})=
(1.20\pm0.14)\times 10^{-4}.
\end{align*}

\begin{figure}[htbp]
   \includegraphics[width=0.5\columnwidth]{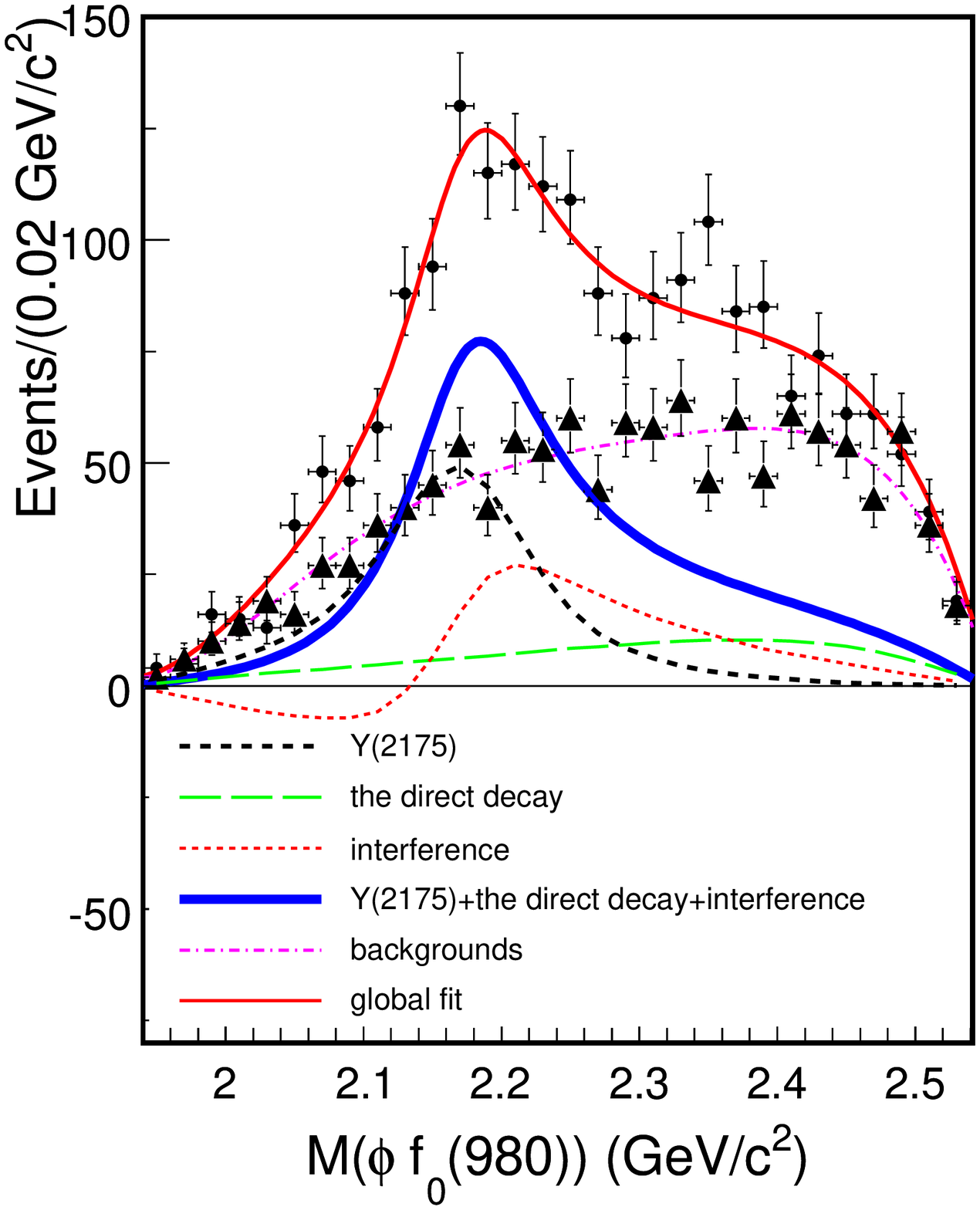}
    \put(-30,120){(a)}
   \includegraphics[width=0.5\columnwidth]{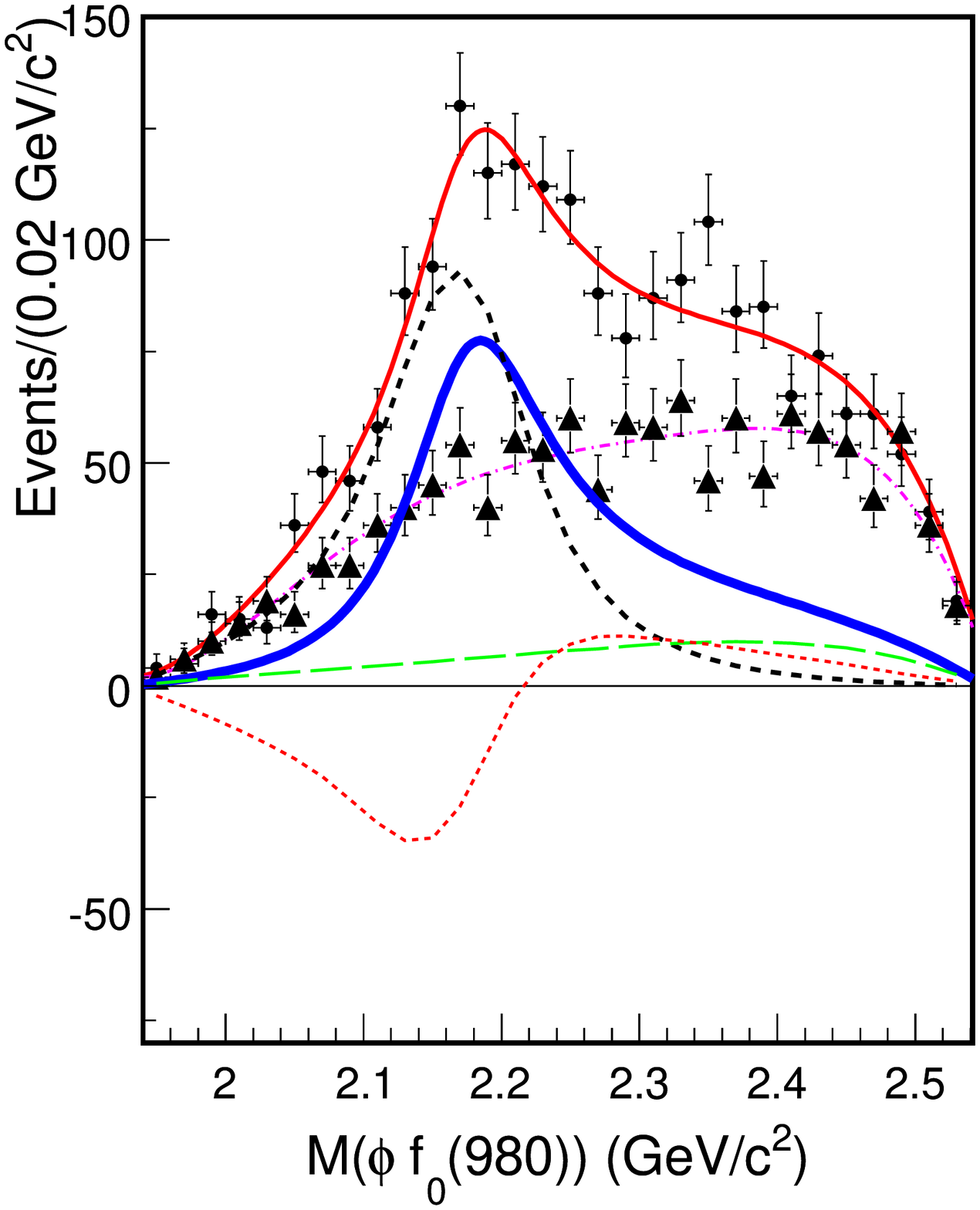}
    \put(-30,120){(b)}
   \caption{The fit projections to the $\phi
       f_0(980)$ invariant mass distribution showing the
       (a)~constructive and (b)~destructive solutions. The
       short-dashed line denotes the signal distribution; the
       dot-dashed curve shows the fit to the backgrounds estimated by
       the sidebands; the long-dashed line denotes the direct decay of
       $J/\psi\to\eta\phi f_0$; and the dotted line denotes the
       interference component.
   \label{doubleso}}
\end{figure}

We also perform a fit to the ${\phi f_0(980)}$ invariant mass,
allowing interference between the $Y(2175)$ and the direct decay
$J/\psi\to\eta\phi f_0(980)$. An ambiguity in the phase angle occurs
when a resonance interferes with a varying
continuum~\cite{bukin}. Thus, two solutions with different relative
phase angles, corresponding to constructive and destructive
interferences, are found. The final fit and the individual
contributions of each of the components are shown in
Fig.~\ref{doubleso}(a),~(b) for constructive and destructive
interference, respectively. The mass, width, and yields of the
$Y(2175)$ signal, as well as the relative phase angle, are shown in
Table~\ref{doublesol}. The statistical significance of the
interference is $2.5\sigma$, which is determined from the differences
of the likelihood values and the degrees of freedom between the fits
with and without interference. In this analysis, the fit results
without considering interference are taken as the nominal values.

\begin {table}[htp]
\caption {Two solutions of the fit to $M(\phi f_0(980))$,
taking interference with the direct decay $\eta \phi f_0$ into account.
Errors are statistical only. }
\label{doublesol}

\begin {tabular}{c c c} \hline
Parameters~& Constructive & Destructive \\  \hline \hline
   M (MeV$/c^2$)                 &$2171\pm10$          &$2170\pm9$      \\  
   $\Gamma$ (MeV)           &$128\pm26$          &$126\pm25$            \\  
    Signal yields            &$400\pm167$              &$744\pm40$      \\  
   relative angle $\Phi$(rad)             &$-0.51\pm0.78$      &$0.60\pm0.64$    \\  \hline
\end {tabular}
\end {table}

\section{\boldmath Measurement of $J/\psi\to\phi f_1(1285)$ and $\phi\eta(1405)$}
\label{X1420_channel}

\begin{figure}[htbp]
   \includegraphics[width=\columnwidth]{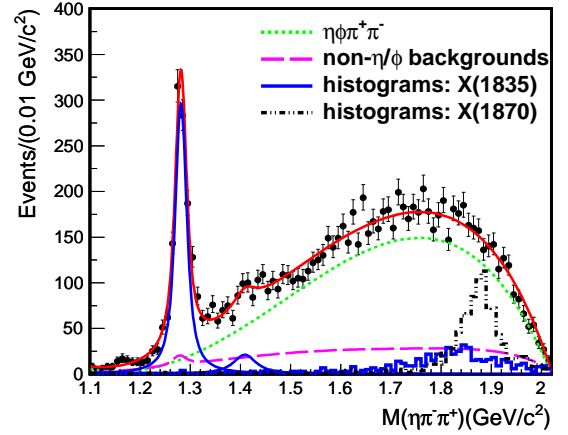}
   \caption{ Fit to the $\eta\pi^{+}\pi^{-}$
       invariant mass spectrum.  The solid lines show the total fit
       and the $f_1(1285)$ and $\eta(1405)$ components; the dashed
       line denotes the non-$\eta$ and non-$\phi$ background estimated
       using the $\eta$-$\phi$ sidebands; the dotted curve represents
       the $J/\psi\to \eta\phi \pi^{+}\pi^{-}$ component; the solid
       histogram indicates the shape of the $X(1835)$ (with arbitrary
       normalization); and the dash-dotted histogram shows the
       predicted shape of the $X(1870)$ signal (with arbitrary
       normalization).
   \label{fit1_up1405}}
\end{figure}

The $\eta\pi^+\pi^-$ mass spectrum recoiling against the $\phi$ is
shown in Fig.~\ref{fit1_up1405}. Besides the significant and
well-known $f_1(1285)$ signal, a small structure around 1.4~GeV$/c^2$,
which is assumed to be the $\eta(1405)$, is evident over a large
non-resonant background. A fit to the ${\eta\pi^+\pi^-}$ invariant mass is
performed with a PDF that includes contributions from the $f_1(1285)$
and $\eta(1405)$ signals, the decay $J/\psi\to\eta\phi\pi^+\pi^-$
(including the process $J/\psi\to\eta\phi f_0(980)$), and backgrounds
from non-$\eta$ and non-$\phi$ processes. In the fit, the $f_1(1285)$
and $\eta(1405)$ signal shapes are described by Breit-Wigner functions
convoluted with Gaussian functions for their mass resolutions. The mass
and width of the $f_1(1285)$ signal are left free in the fit, while those of the
$\eta(1405)$ signal are fixed to the values in the PDG~\cite{PDG}. The
parameters of the Gaussian functions for the mass resolutions are
fixed to their MC values. The shape of the
$J/\psi\to\eta\phi\pi^+\pi^-$ decay is represented by a third-order
Chebychev polynomial function, and the corresponding parameters are
allowed to vary. The non-$\eta$ and non-$\phi$ background is estimated with
the events in the $\eta$-$\phi$ sideband regions, as shown by the
dashed lines in Fig.~\ref{fit1_up1405}, and is fixed in the fit.

The fit, shown in Fig.~\ref{fit1_up1405}, yields $1154\pm56$
$f_1(1285)$ signal events, with a mass of $1281.7\pm0.6$~MeV$/c^{2}$
and a width of $21.0\pm1.7$~MeV. The mass and width are in good
agreement with world average values~\cite{PDG}. Using a detection
efficiency of $(22.14\pm0.09)\%$, obtained from MC simulation, the
product branching fraction is measured to be:
\begin{align*}
 \mathcal{B}(&J/\psi\to\phi f_1(1285), \\
 & f_1(1285)\to\eta\pi^{+}\pi^{-}) =(1.20\pm0.06)\times 10^{-4},
\end{align*}
where the error is statistical only.

For the $\eta(1405)$ signal, the fit yields $172\pm50$ events with a
statistical significance of $3.6\sigma$, evaluated from the difference
of the likelihood values between the fits with and without the
$\eta(1405)$ included. The product branching fraction is
$\mathcal{B}(J/\psi\to\phi
\eta(1405),\eta(1405)\to\eta\pi^{+}\pi^{-})$=$(2.01\pm0.58)\times
10^{-5}$, where the error is statistical only. To determine the upper
limit on the $\eta(1405)$ production rate, a series of similar fits
with given numbers of $\eta(1405)$ events are performed, and the
likelihood values of the fits as a function of the number of
$\eta(1405)$ events are taken as a normalized probability
function. The upper limit on the number of signal events at the $90\%$
C.L., $N^{U.L.}$, is defined as the value that contains $90\%$ of the
integral of the normalized probability function. The fit-related
uncertainties on $N^{U.L.}$ are estimated by using different sideband
regions for the effect of the non-$\eta$ and non-$\phi$ background,
different orders of Chebychev polynomials for the shape of the
$J/\psi\to \eta\phi \pi^{+}\pi^{-}$ and changing the mass and width
values of the $\eta(1405)$ within one standard deviation from the
central values for the signal shape. Finally, after taking into
account fit-related uncertainties, we obtain $N^{U.L.}$ = $345$. This upper
limit and the detection efficiency of $(19.75\pm0.12)\%$, estimated from
MC simulation, are used to evaluate the upper limit on the branching
fraction:
\begin {equation}
\label{eqfit2}
\begin{split}
&\mathcal{B}(J/\psi\to\phi \eta(1405), \eta(1405)\to\eta\pi^{+}\pi^{-}))\\
&<\frac{N^{U.L.}}{\epsilon\times N_{J/\psi}\times
\mathcal{B}(\eta\to\gamma\gamma)\times \mathcal{B}(\phi\to
K^{+}K^{-}) \times (1-\sigma_\mathrm{sys})}\\
&=4.45\times10^{-5},\\
\end{split}
\end {equation}
where $\sigma_\mathrm{sys}$ is the systematic error to be discussed in detail
below.  Since the background uncertainty is taken into account in the
calculation of $N^{U.L.}$ by choosing the maximum event yield from the
variations of the background functions, the systematic uncertainty from this
source is excluded here. The final results on the upper limit of the
branching fraction are shown in Table~\ref{sumres2}.

In the $\eta\pi^+\pi^-$ mass spectrum shown in Fig.~\ref{fit1_up1405},
we do not observe obvious structures around 1.84~GeV/c$^2$ or at
1.87~GeV/c$^2$. Using the same approach as was used for the
$\eta(1405)$, we set 90\%~C.L. upper limits for the $X(1835)$ and
$X(1870)$ production rates, where the signal shape of the $X(1835)$ or
$X(1870)$ is described by a Breit-Wigner function convoluted with a
Gaussian function for the mass resolution, and the background is
modeled by a third-order Chebychev polynomial. The resonant parameters
of the $X(1835)$ and $X(1870)$ are fixed to the values of previous
BESIII measurements~\cite{X1835_pipietap2,liuk}. The results are
summarized in Table~\ref{sumres1} and Table~\ref{sumres2}.

\begin {table}[htp]
\begin {center}
\caption { Measurements of the number of events, statistical significances, and efficiencies.
}
\label{sumres1}
\begin {tabular}{cccc} \hline
Resonance &$N_{obs}$&Significance&Efficiency($\%$)
\\  \hline \hline
    $Y(2175)$              &$471\pm54$     &$>10\sigma$     &$9.10\pm0.01$      \\
    $f_{1}(1285)$              &$1154\pm56$       &$-$     &$22.14\pm0.09$       \\
    $\eta(1405)$              &$172\pm50(<345)$       &$3.6\sigma$     &$19.75\pm0.12$       \\
    $X(1835)$                 &$394\pm360(<1522)$     &$1.1\sigma$     &$13.85\pm0.14$       \\
    $X(1870)$                 &$25\pm73(<330)$       &$0.8\sigma$     &$13.73\pm0.14$         \\ \hline
\end {tabular}
\end {center}
\end {table}

\begin {table*}[htp]
\begin {center}
\caption { Measurements of the branching fractions for the decay modes.  Upper limits are given at the $90\%$ C.L.}
\label{sumres2}
\begin {tabular}{cc} \hline
Decay mode &Branching fraction $\mathcal{B}$
\\  \hline \hline
$J/\psi\to\eta Y(2175)$, $Y(2175)\to\phi f_0(980)$, $f_0(980)\to\pi^{+}\pi^{-}$                       &$(1.20\pm0.14\pm0.37)\times 10^{-4}$    \\
$J/\psi\to\phi f_1(1285)$, $f_1(1285)\to\eta\pi^{+}\pi^{-}$                                          &$(1.20\pm0.06\pm0.14)\times 10^{-4}$   \\
$J/\psi\to\phi \eta(1405)$, $\eta(1405)\to\eta\pi^{+}\pi^{-}$                                        &$(2.01\pm0.58\pm0.82)(<4.45)\times 10^{-5}$   \\
$J/\psi\to\phi X(1835)$, $X(1835)\to\eta\pi^{+}\pi^{-}$                                                &$<2.80\times 10^{-4}$ \\
$J/\psi\to\phi X(1870)$, $X(1870)\to\eta\pi^{+}\pi^{-}$                                                    &$<6.13\times 10^{-5}$\\  \hline
\end {tabular}
\end {center}
\end {table*}

\section {Systematic errors}

The sources of systematic error include: the efficiency difference
between data and MC simulation for the track reconstruction, the PID,
the photon detection, and the kinematic fit; the fitting procedure;
the ambiguity in the interference; and the number of $J/\psi$
events. Their effects on the measurement of the resonance parameters
and the branching fractions are discussed in detail below.

\emph{a. MDC Tracking efficiency~~} The tracking efficiency has been
investigated using the almost background-free control samples of
$J/\psi\to\pi^{+}\pi^{-}p\overline{p}$ and $J/\psi\to K_{S}^0 K \pi$
~\cite{trackerror}. The difference in tracking efficiency between data
and MC is found to be 2\% per charged kaon and pion. Therefore, 8\% is taken as
the total systematic error for the detection efficiency of four
charged tracks.

\emph{b. PID efficiency~~} To evaluate the PID efficiency uncertainty,
we have studied the kaon and pion PID efficiencies using the control
samples of $J/\psi\to K^{*\pm} K^\mp$ and
$J/\psi\to\rho\pi$~\cite{trackerror}, respectively. The difference in
PID efficiency between data and MC is 1\% per kaon and pion. Hence,
4\% is taken as the total systematic error from the PID efficiency.

\emph{c. Photon detection efficiency~~} The photon detection
efficiency has been studied using a control sample of
$J/\psi\to\rho\pi$~\cite{trackerror}. The difference between data and
MC is found to be 1\% per photon. Therefore, 2\% is taken as the total
systematic error for the efficiency of the detection of the two
photons.

\emph{d. Kinematic fit~~} To estimate the uncertainty associated with
the kinematic fit, a control sample of
$J/\psi\to\phi\eta^{\prime}(\eta^\prime\rightarrow \eta\pi^+\pi^-)$,
which has exactly the same final state as the signal, is first
selected without a kinematic fit. The kinematic fit efficiency is then
evaluated from the ratio of the $\eta^{\prime}$ yields with and
without the kinematic fit requirement, where the $\eta^{\prime}$ yield
is extracted from the fit to the $\eta^\prime$ signal in the
$\eta\pi^+\pi^-$ invariant mass. The difference of the kinematic fit
efficiency between data and MC, $0.4\%$, is taken as the systematic
error for the kinematic fit.

\emph{e. Uncertainties of $\mathcal{B}(\eta\rightarrow\gamma\gamma)$
  and $\mathcal{B}(\phi\rightarrow K^+K^-)$~~} The branching fractions
of $\eta\rightarrow\gamma\gamma$ and $\phi\rightarrow K^+K^-$ are
taken from the PDG~\cite{PDG}. The uncertainties of these branching
fractions, $0.5\%$ and $1.0\%$, are taken as the systematic errors.

\emph{f. Uncertainty of the number of $J/\psi$ events~~} The total
number of $J/\psi$ events is determined from an analysis of inclusive
$J/\psi$ hadronic decays, and the uncertainty of the number of
$J/\psi$ events, $1.2\%$~\cite{jpsiN}, is taken as the systematic
error from the number of $J/\psi$ events.

\emph{g. Background uncertainty~~} In the measurement of the resonance
parameters and branching fractions of the $Y(2175)$, a fit is
performed to the $\phi f_0(980)$ invariant mass spectrum. In the fit,
the shape and amplitude of the background from the non-$\phi$ and
non-$f_0(980)$ are fixed to the estimation from the $\phi-f_0(980)$
sideband regions. To estimate its impact on the final results, we use
different $\phi-f_0(980)$ sideband regions to estimate the background,
and follow the same fit procedure. The maximum changes on both the
$Y(2175)$ resonance parameters and its signal yield are taken as
the systematic errors. The uncertainty due to the background on the
mass and width of the $Y(2175)$ are $\pm4.0$~MeV$/c^2$ and
$\pm14.0$~MeV, respectively.

For the branching fraction of $J/\psi\to\phi f_1(1285)/\eta(1405)$
with $f_1(1285)/\eta(1405)\to\eta\pi^{+}\pi^{-}$, the non-$\eta$ and
non-$\phi$ backgrounds are estimated with the events in the
$\eta-\phi$ sideband regions. Analogous to the evaluation of the
$Y(2175)$ errors, we define different sideband regions to estimate
the backgrounds and follow the same fit procedure. The largest
changes are taken as the uncertainty from the background for these
measurements. Compared to the number of $f_1(1285)$ events, the
fluctuation of background shape under the $\eta(1405)$ peak has a
large impact on the signal yields in the fit due to the limited
statistics.

\emph{h. Impact from possible extra resonances~~~} In the invariant
mass spectrum of $\phi f_{0}(980)$, a small structure around 2.35~GeV/$c^2$
is found (Fig.~\ref{fitY2}).  To evaluate its impact on the $Y(2175)$
measurement, we perform a fit with an additional signal around
$2.35$~GeV/$c^2$, which is described with a Breit-Wigner function
convoluted with a Gaussian function for the mass resolution. The fit
results show that the significance of the structure around
2.35~GeV/c$^2$ is only 3.8$\sigma$.  It is therefore not considered in
the nominal final results. However, the impact on the $Y(2175)$ measurement is
taken as the systematic error. The uncertainty due to the possible
extra resonance on the mass and width of the $Y(2175)$ are
$\pm3.0$~MeV$/c^2$ and $\pm5.0$~MeV, respectively.

In the measurement of the branching fraction of $J/\psi\to\phi
f_1(1285)$ with $f_1(1285)\to\eta\pi^{+}\pi^{-}$, we perform a fit
without the $\eta(1405)$ signal. The difference of results with or
without the $\eta(1405)$ signal included in the fit is taken as the
systematic error on the $f_1(1285)$ measurement from the impact of the
$\eta(1405)$.

\emph{i. Parameterization of the $f_0(980)$~~} The systematic error
from the $f_{0}(980)$ shape is estimated by comparing the detection
efficiencies from the signal MC samples simulated with different
parameterizations of the $f_{0}(980)$. We use the resonant parameters
of the $f_{0}(980)$ from Ref.~\cite{otherpara}, instead of the nominal
values from the measurements of BESII~\cite{phipipi} mentioned in
Section II, to describe the $f_{0}(980)$ shape. This leads to a
difference in the detection efficiency of $7.6\%$, and is taken as the
systematic uncertainty on the $Y(2175)$ branching fraction measurement
from the $f_0(980)$ parameterization.

\emph{j. Uncertainty from fixed mass and width values on the branching
  ratio of $J/\psi\to\phi\eta(1405)$ with
  $\eta(1405)\to\eta\pi^{+}\pi^{-}$~~} The mass and width of the
$\eta(1405)$ are fixed to their PDG values in the fit to the
$\eta(1405)$ signal. We change the mass and width values by one standard
deviation from their central values in the fitting procedure. The
maximum change on the branching fraction is determined to be $7.0\%$
when the mass and width values are fixed at one negative standard
deviation from the central values.

\emph{k. Uncertainty from parameter sets in the generation of
  $J/\psi\to\phi f_1(1285)$~~} The parameters used in the generation
of the signal MC sample of $J/\psi\to\phi f_1(1285)$ are taken from the
angular distribution of the $\phi$ in the rest frame of the $J/\psi$
found in real data. The impact of the uncertainty of these parameters on the
efficiency, $3.2\%$, is taken as a source of systematic error on the
branching fraction.

\begin {table*}[htp]
\caption{Summary of systematic errors (in \%) for the branching
  fraction measurements. The fourth column shows the sources of
  systematic errors on the branching fraction of
  $J/\psi\to\phi\eta(1405)$ with $\eta(1405)\to\eta\pi^{+}\pi^{-}$,
  while the fifth column shows those on the upper limits of the
  branching fractions of $J/\psi\to\phi\eta(1405)$, $\phi X(1835)$,
  $\phi X(1870)$ with
  $\eta(1405)/X(1835)/X(1870)\to\eta\pi^{+}\pi^{-}$. }\label{totalerror_br}
\begin{small}
\begin {tabular}{c c c c c} \hline
    Sources                      &~~$Y(2175)$~~&~~$f_1(1285)$~~&~~$\eta(1405)$~~&~~$\eta(1405)/X(1835)/X(1870)$~~   \\  \hline \hline
    MDC tracking                 &\multicolumn{4}{c}{$8.0$}               \\
    Photon detection             &\multicolumn{4}{c}{$2.0$}                 \\
    PID                          &\multicolumn{4}{c}{$4.0$}                 \\
    Kinematic fit                &\multicolumn{4}{c}{$0.4$}                 \\
    $\mathcal{B}(\eta\to\gamma\gamma)$    &\multicolumn{4}{c}{$0.5$}                 \\
    $\mathcal{B}(\phi\to K^{+}K^{-})$    &\multicolumn{4}{c}{$1.0$}                  \\
    Number of $J/\psi$ events       &\multicolumn{4}{c}{$1.2$}                  \\
    $f_{0}(980)$ selection       &$7.6$       &$-$         &$-$      &$-$          \\
    Background uncertainty       &$19.1$      &$4.1$       &$39.3$   &$-$          \\
    The fixed $M/\Gamma$ of $\eta(1405)$ &$-$  &$-$ &$7.0$   &$-$          \\
    Parameters of $\phi f_1(1285)$ generation &$-$ &$3.2$ &$-$ &$-$            \\
    Extra resonance              &$21.4$      &$4.0$       &$-$      &$-$          \\  \hline
    Total                        &$31.1$      &$11.4$      &$41.0$    &$9.4$    \\  \hline
\end {tabular}
\end{small}
\end {table*}

In Table~\ref{totalerror_br}, a summary of all contributions to the
systematic errors on the branching fraction measurements is shown. In
each case, the total systematic uncertainty is obtained by adding the
individual contributions in quadrature. For the uncertainties on the
$Y(2175)$ resonant parameters, we find that the dominant systematic
uncertainties are from the background shape and a possible additional
resonance around 2.35~GeV/$c^2$. Adding the various systematic
uncertainties in quadrature, the total systematic errors on the mass
and width of the $Y(2175)$ are $\pm 5.0$~MeV$/c^2$ and $\pm 14.8$~MeV,
respectively.

\section{Summary}

In summary, we present an analysis of $J/\psi\to\eta \phi
\pi^{+}\pi^{-}$  based on $(225.3\pm2.8)\times 10^{6}$ $J/\psi$ events
collected with the BESIII detector.   The $Y(2175)$ resonance is
observed in the invariant mass spectrum of $\phi f_{0}(980)$ with a
statistical significance of greater than $10\sigma$. The mass and width of the
$Y(2175)$ are measured and are in good agreement with previous
experimental results~(Table~\ref{3res}). Neglecting the effects of
interference with the direct decay  $J/\psi\to\eta\phi f_{0}(980)$,
the product branching fraction is measured to be
$\mathcal{B}(J/\psi\to\eta Y(2175)$, $Y(2175)\to\phi f_{0}(980)$,
$f_{0}(980)\to\pi^{+}\pi^{-})=(1.20\pm0.14\pm0.37)\times10^{-4}$. We
also perform a fit taking the interference between
the $Y(2175)$ and the direct decay. The
corresponding results are shown in Table~\ref{doublesol}.
\begin {table*}[htp]
\caption { Comparison of $Y(2175)$ parameters as measured by different experiments.}
\label{3res}
\begin {tabular}{c c c c} \hline
 Collaboration&Process&$M$ (MeV$/c^{2}$) &$\Gamma$ (MeV) \\ \hline\hline
     BABAR~\cite{aaaa_babar}              &$e^{+}e^{-}\to\phi f_{0}~~(ISR)$     &$2175\pm10\pm15$   &$58\pm16\pm20$ \\      
     BESII~\cite{aaab_wanx}              &$J/\psi\to\eta \phi f_{0}(980)$   &$2186\pm10\pm6$    &$65\pm23\pm17$                              \\      
     BELLE~\cite{aaba_belle}              &$e^{+}e^{-}\to\phi f_{0}~~(ISR)$     &$2079\pm13^{+79}_{-28}$   &$192\pm23^{+25}_{-61}$             \\      
     BABAR(updated)~\cite{babar_y2175}     &$e^{+}e^{-}\to\phi f_{0}~~(ISR)$     &$2172\pm10\pm8$    &$96\pm19\pm12$                      \\      
     BESIII             &$J/\psi\to\eta \phi f_{0}(980)$    &$2200\pm6\pm5$   &$104\pm15\pm15$                             \\ \hline
\end {tabular}
\end {table*}

In addition, we investigate the $\eta \pi^{+}\pi^{-}$ mass spectrum
recoiling against the $\phi$ in the $J/\psi$ decay. A structure
around 1.28~GeV$/c^2$ is clearly seen, and the fit results are in good
agreement with the world average values of the $f_1(1285)$
parameters. The product branching fraction of $J/\psi\to\phi
f_1(1285)$ with $f_1(1285)\to\eta\pi^{+}\pi^{-}$ is measured to be
$\mathcal{B}(J/\psi\to\phi f_1(1285)\to
\phi\eta\pi^{+}\pi^{-})=(1.20\pm0.06\pm0.14)\times10^{-4}$. A structure
around $1.4$~GeV$/c^2$ seems to be present in the $\eta\pi^{+}\pi^{-}$
mass spectrum. Assuming it to be the $\eta(1405)$, the product branching
fraction is calculated to be $\mathcal{B}(J/\psi\to\phi
\eta(1405)\to\phi\eta\pi^{+}\pi^{-})=(2.01\pm0.58\pm0.82)\times10^{-5}$.
We also present a $90\%$~C.L. upper limit on the branching fraction
$\mathcal{B}(J/\psi\to\phi
\eta(1405)$,$\eta(1405)\to\eta\pi^{+}\pi^{-})<4.45\times 10^{-5}$. In
a previous experiment, the $\eta(1405)/\eta(1440)$ is observed in both
$\eta\pi\pi$ and $K\overline{K}\pi$ invariant mass spectra recoiling
against the $\gamma$ and $\omega$ in $J/\psi$ decays. However, no
significant structure around 1.4~GeV$/c^2$ is observed in the
$\pi^+\pi^-\eta$ mass spectrum recoiling against the $\phi$ in this
analysis, which may imply that $u$ and $d$ quarks account for more of the
quark content in the $\eta(1405)$ than the $s$ quark. We also
perform searches for the $X(1835)$ and $X(1870)$ in the vicinity of
1.8~GeV$/c^2$ in the $\eta\pi^{+}\pi^{-}$ mass spectrum, and observe
no evident structures.  The corresponding upper limits at
$90\%$~C.L. of branching fraction are measured. All of these
measurements provide information in understanding the nature of the
$X(1835)$ and $X(1870)$.

\section*{Acknowledgements}
The BESIII collaboration thanks the staff of BEPCII and the IHEP
computing center for their strong support. This work is supported in
part by National Key Basic Research Program of China under Contract
No.~2015CB856700; Joint Funds of the National Natural Science
Foundation of China under Contracts Nos.~11079008, 11179007, U1232201,
U1332201; National Natural Science Foundation of China (NSFC) under
Contracts Nos.~10935007, 11121092, 11125525, 11235011, 11322544,
11335008, 11175189; the Chinese Academy of Sciences (CAS) Large-Scale Scientific
Facility Program; CAS under Contracts Nos.~KJCX2-YW-N29, KJCX2-YW-N45;
100 Talents Program of CAS; INPAC and Shanghai Key Laboratory for
Particle Physics and Cosmology; German Research Foundation DFG under
Contract No.~Collaborative Research Center CRC-1044; Istituto
Nazionale di Fisica Nucleare, Italy; Ministry of Development of Turkey
under Contract No.~DPT2006K-120470; Russian Foundation for Basic
Research under Contract No.~14-07-91152; U.S.\ Department of Energy
under Contracts Nos.~DE-FG02-04ER41291, DE-FG02-05ER41374,
DE-FG02-94ER40823, DESC0010118; U.S.\ National Science Foundation;
University of Groningen (RuG) and the Helmholtzzentrum fuer
Schwerionenforschung GmbH (GSI), Darmstadt; WCU Program of National
Research Foundation of Korea under Contract No.~R32-2008-000-10155-0.

\end{document}